**Publication boost in Web of Science journals and its effect on citation distributions**

Lovro Šubelj[a, *]
Dalibor Fiala[b]

[a] University of Ljubljana, Faculty of Computer and Information Science
Večna pot 113, 1000 Ljubljana, Slovenia

[b] University of West Bohemia, Department of Computer Science and Engineering
Univerzitní 8, 30614 Plzeň, Czech Republic

* Corresponding author. Tel.: +386 (0)1 479 82 33.
Email addresses: lovro.subelj@fri.uni-lj.si (L. Šubelj), dalfia@kiv.zcu.cz (D. Fiala).

**Abstract**: In this paper we show that the dramatic increase in the number of research articles indexed in the Web of Science database impacts the commonly observed distributions of citations within these articles. First, we document that the growing number of physics articles in recent years is due to existing journals publishing more and more papers rather than more new journals coming into being as it happens in computer science. And second, even though the references from the more recent papers generally cover a longer time span, the newer papers are cited more frequently than the older ones if the uneven paper growth is not corrected for. Nevertheless, despite this change in the distribution of citations, the citation behavior of scientists does not seem to have changed.

**Introduction**

It is well known that scientific communication has changed dramatically in recent decades. There has been a real publication boom with more and more papers published, indexed in databases, available online, and cited. All of this might have had some impact on the way research papers refer to each other and citation patterns come into existence. Although not necessarily all newly indexed publications in bibliographic databases are the result of new research (Michels and Schmoch (2012) showed that half of the growth in the number of papers indexed in Web of Science from 2000 to 2008 was caused by the inclusion of previously existing journals), the growth of scientific production is undeniable and the question is whether this growth is accompanied by some novel trends in the citation patterns of research papers. In this study, we investigate this issue by analyzing two large Web of Science data sets consisting of computer science and physics journal articles, and conclude that the enormous increase in research publications alters commonly observed citation distributions. Nevertheless, when this growth is corrected for, the citation behavior of scientists appears not to change.

Citation patterns of research papers and their change over the course of time were the concern of many previous studies. For instance, Egghe (2010) introduced a mathematical model of the aging of references. Larivière et al. (2009) documented that the age of cited references declined between 1900 and 2005, but, in contrast, Verstak et al. (2014) showed that more and more older papers are cited in current literature. As far as citation models are concerned, Eom and Fortunato (2011) modeled citation distributions in the papers from the American Physical Society (APS) journals and discovered the shifted power law function to best describe the citation patterns in the network under study. A combined exponential and power law citation model was proposed by Peterson et al. (2010). Newman (2014) describes a successful method for predicting the future impact of articles and another impact prediction model is discussed by Wang et al. (2013). Unlike the latter, Stegehuis et al. (2015) proposed a model that can be adopted soon after the publication of an article. Radicchi and Castellano (2011) inspected the citation distributions of all articles in APS journals between 1985 and 2009 in individual years and fields, and proposed rescaling factors that would enable impartial comparisons of citedness. A similar procedure was conducted by Radicchi et al. (2008) for papers from 20 different research disciplines. Also, Radicchi et al. (2009) analyzed the whole collection of Physical Review papers (i.e. APS) from 1893 to 2006 and proposed a diffusion algorithm of scientific credit to rank authors by importance. Another study of this sort was that by Walker et al. (2007). Note that some of the above studies considered citation distributions, i.e. the proportion of citations papers published in a certain year receive in the subsequent years, whereas others focused on the proportion of papers that receive a certain number of citations over the same time period, i.e. the in-degree distributions of citation networks. Nevertheless, as we show below, our findings impact both types of studies.

**Data and Methods**

In late 2014, we generated two citation networks of research articles: Computer Science (Web of Science papers categorized as "computer science") with 492,124 nodes (articles) and 2,328,599 edges (citations) and Physics (Web of Science papers categorized as "physics") with 1,793,665 articles and 20,299,195 citations. We include merely journal articles and reviews, and discard all notes, letters, corrections, meeting abstracts, proceedings papers, book reviews and other. Also, in our data sets there are no citations from other fields and we do not study references to publications from other fields. We investigate only the in-field citations within computer science on one hand and within physics on the other. Note, however, that the two research areas are not mutually exclusive and there are papers belonging to both of them. Both data sets span from the beginning of Web of Science until 2014 and were selected as they show quantitatively different behavior. Particularly, we analyzed the two data sets in terms of publication and journal counts, citation and reference distributions in various years, and in-degree power law exponents of citation distributions in 10-year intervals, and obtained results that are discussed in the next section.

**Results and Discussion**

The production growth in both scientific disciplines was exponential in the time period under study as we may see in Figure 1, where the publication counts at 5-year intervals from 1975 to

2010 are shown in the plot on the left-hand side. The production increase accelerated towards the end of the time span (in 2005 and 2010 in physics with more than 50,000 articles and in 2010 in computer science with nearly 30,000 articles) with the growth rate in computer science being much higher than that in physics. A completely different picture can be seen on the right-hand side of Figure 1 where the linear increase in journal counts for computer science and physics is depicted over time. While the number of physics journals indexed in Web of Science increased only moderately from about 100 to roughly 150 in 35 years (with even a small decline between 2000 and 2005), the number of computer science journals exploded from about 50 in 1975 to almost 450 in 2010. The dynamics of this growth was at its top between 1990 and 1995 when the journal count almost doubled. Therefore, by comparing the two plots in Figure 1, it seems that the increase in the number of computer science articles is triggered by the massive growth in the number of computer science journals, but the growing amount of physics articles is rather caused by more frequent issues or bigger volumes of the existing journals.

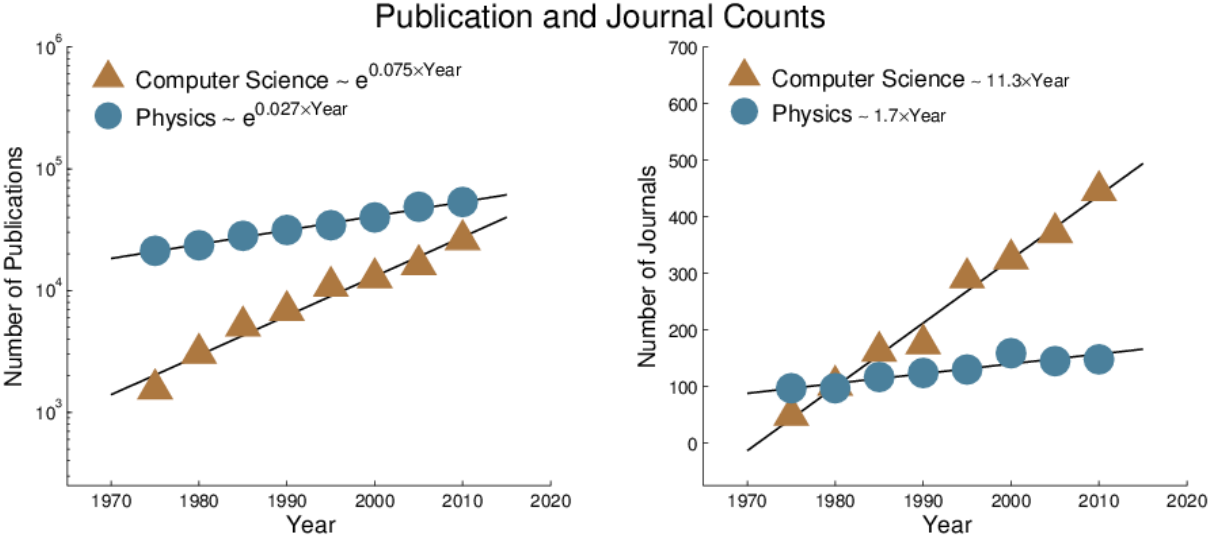

**FIG. 1.** Publication and journal counts in different years (straight lines show the least-squares fits to a linear or exponential function).

As far as the distribution of references to other papers in the articles under study is concerned, we refer to the plots in Figure 2. For computer science in the top-left plot, we can see the curves depicting the proportion of cited papers, published in individual years, from articles published between 1985 and 2010 in 5-year spans. (Thus, we omitted the first year in our data set, 1975, because parsing the references from that year's articles would actually have meant expanding the data set beyond the data at hand.) We may notice that although the reference peak always occurs roughly two years before the citing article is published, i.e. the most references refer to 2-year old papers, it generally becomes lower and the tail of the reference curve gets longer and thus less steep over the whole time period. For instance, almost 16% of references from the articles published in 1980 refer to 2-year old papers, but it is less than 10% for the articles from 2010. It also seems that about 5% of references from 1980 articles cite 5-year old papers, but the same proportion of references from 2010 articles are even made to 10-year old publications. Thus, more recent articles tend to cite older papers to a greater extent than it was common in the past, an effect already observed by Larivière et al. (2009) and more recently by Verstak et al. (2014). The same roughly holds for physics, in the bottom-left plot of Figure 2, with the notable difference that the reference peaks remain stable with about 13% up until 1990 and then only moderately decline in later years. We may speculate that the recent tendency of papers to cite older publications is responsible for the emergence of novel "sleeping beauties" in science as discovered by Ke et al. (2015), whereas further research would be needed to verify this claim. In addition to the left-hand plots, we show the curves collapsed one on the top of the other using the transformed variable "Cited Year – Peak Year" in the right-hand plots of Figure 2. There we can clearly see that the reference distributions did not change over the past decades and, moreover, the distance to the peak year denoted "Peak Year Delta" remained stable in physics and slightly increased in computer science as depicted in the small inset plots.

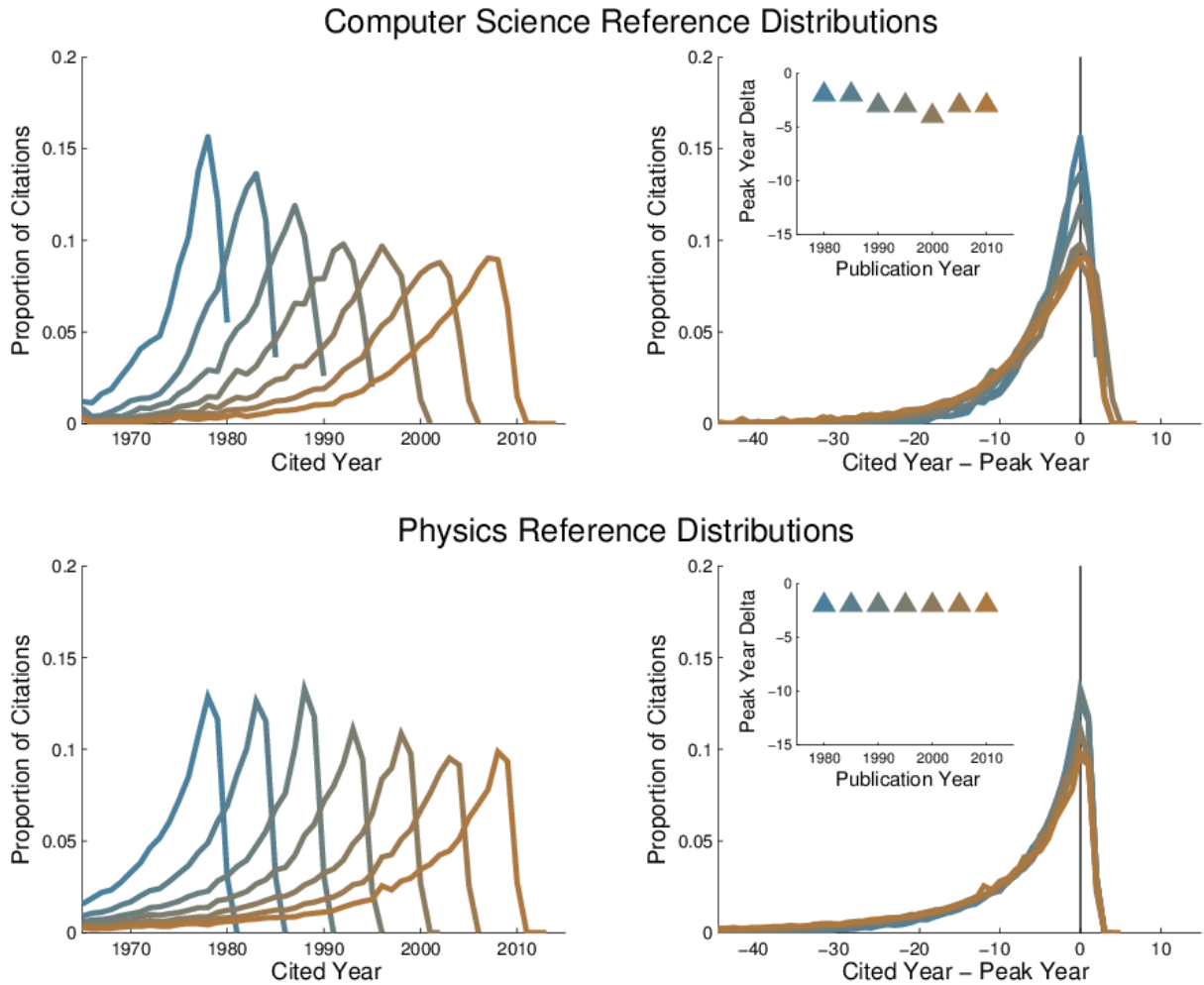

**FIG. 2.**   Reference distributions in different years.

By analogy, we also consider distributions in the opposite direction, from the cited articles to the citing ones. More precisely, denote $n_y$ to be the number of papers published in year $y$ and $n_y^x$ the number of citations from papers in year $x$ to papers in year $y$, $x \geq y - 1$. ($-1$ here is due to the fact that some papers receive citations even before publication as discussed below.) Then, the citation distribution of papers in year $y$ is defined as $P_y(x) = n_y^x / \sum_{\hat{x} \geq y-1} n_y^{\hat{x}}$. Looking at the citation distributions in Figure 3, we may immediately notice a substantial difference between computer science and physics. While in physics (bottom-left plot) the sharp citation peaks occurring some two years after publication slightly increase over the time period under study from 9% to 14% of citations between 1975 and 2005 (with 2010 being left out for similar reasons as 1975 in the analysis of references) followed by smooth long tails.

On the other hand, the citation peaks are by far not so sharp in computer science (top-left plot of Figure 2), but even quite broad for the articles published in 2000 and 2005 and increasing from 4% to 14% of citations for the articles at the beginning and at the end of the time span investigated. In addition, the tails of the citation curves are not smooth but rather rugged. Also, the peak year distances are quite variable in computer science compared to the stability of physics (see the inset plots on the right-hand side of Figure 3).

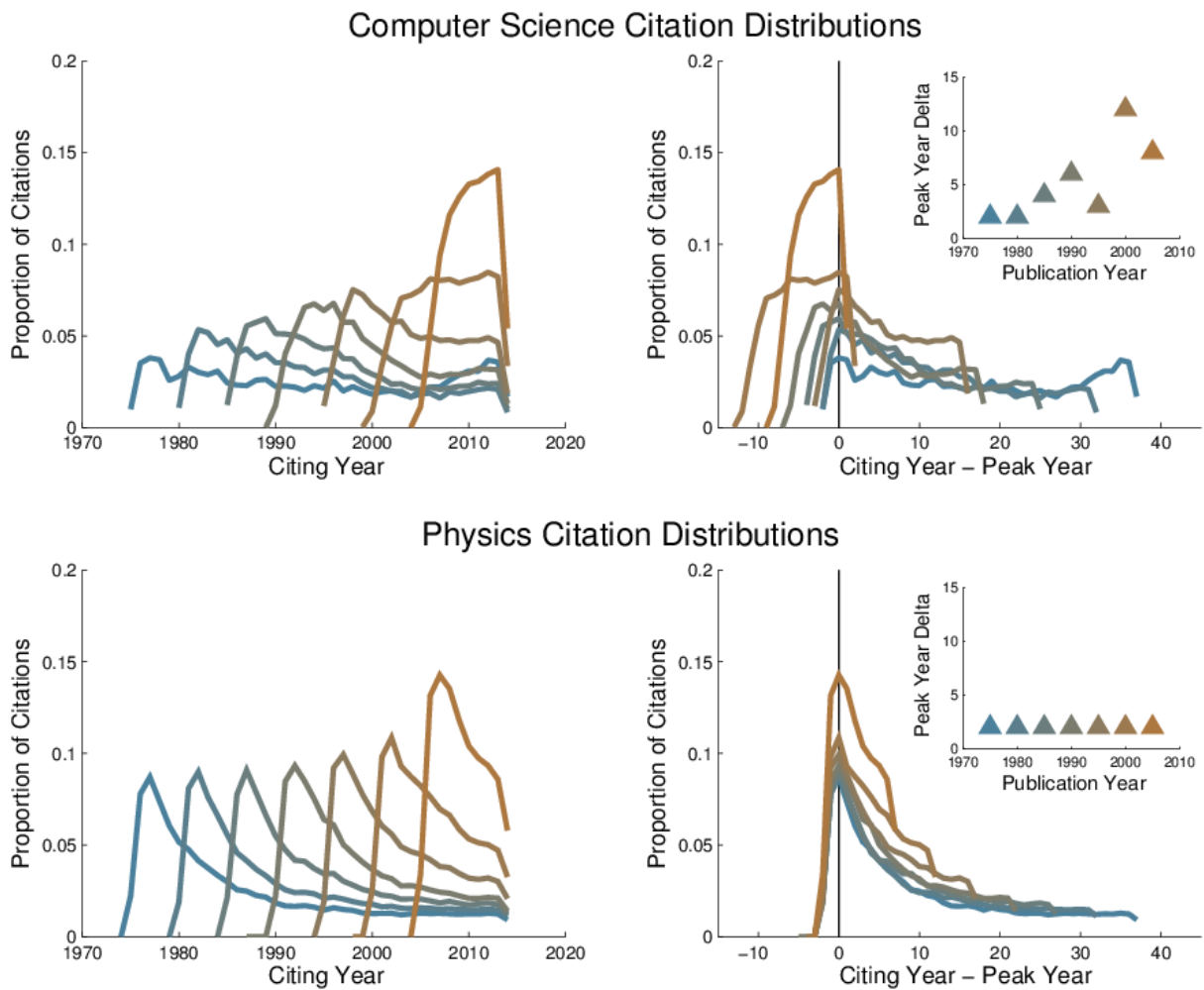

**FIG. 3.**  Citation distributions in different years.

So, do the broad peaks, rough tails, and varying peak year distances mean a different and changing citation behavior in computer science? In fact, not really, as we can easily see in Figure 4. Here the citations are normalized, and thus divided, by the number of published articles in each citing year to correct for the different amounts of publications in the individual years. Hence, we define normalized citation distributions of papers in year $y$ as $\widehat{P}_y(x) = (n_y^x/n_x) / \sum_{\hat{x} \geq y-1}(n_y^{\hat{x}}/n_{\hat{x}})$. While the curve shapes remain almost unchanged in physics (bottom-left plot of Figure 4), we may see that they changed dramatically in computer science (in the top-left plot) and now resemble the normalized citation distribution curves in physics to a great extent with only the citation peaks for the 1975 and 1980 curves being somewhat higher. Even the peak year distances (in the upper inset plot) are now closer to those in physics (in the lower inset). Therefore, we may draw the conclusion here that the citation behavior changed neither in physics nor in computer science over the years when citation counts are properly normalized to reflect the growing number of publications. Furthermore, the effect of normalization is much more visible in computer science than in physics due to the much stronger publication growth in recent years (see Figure 1 for evidence).

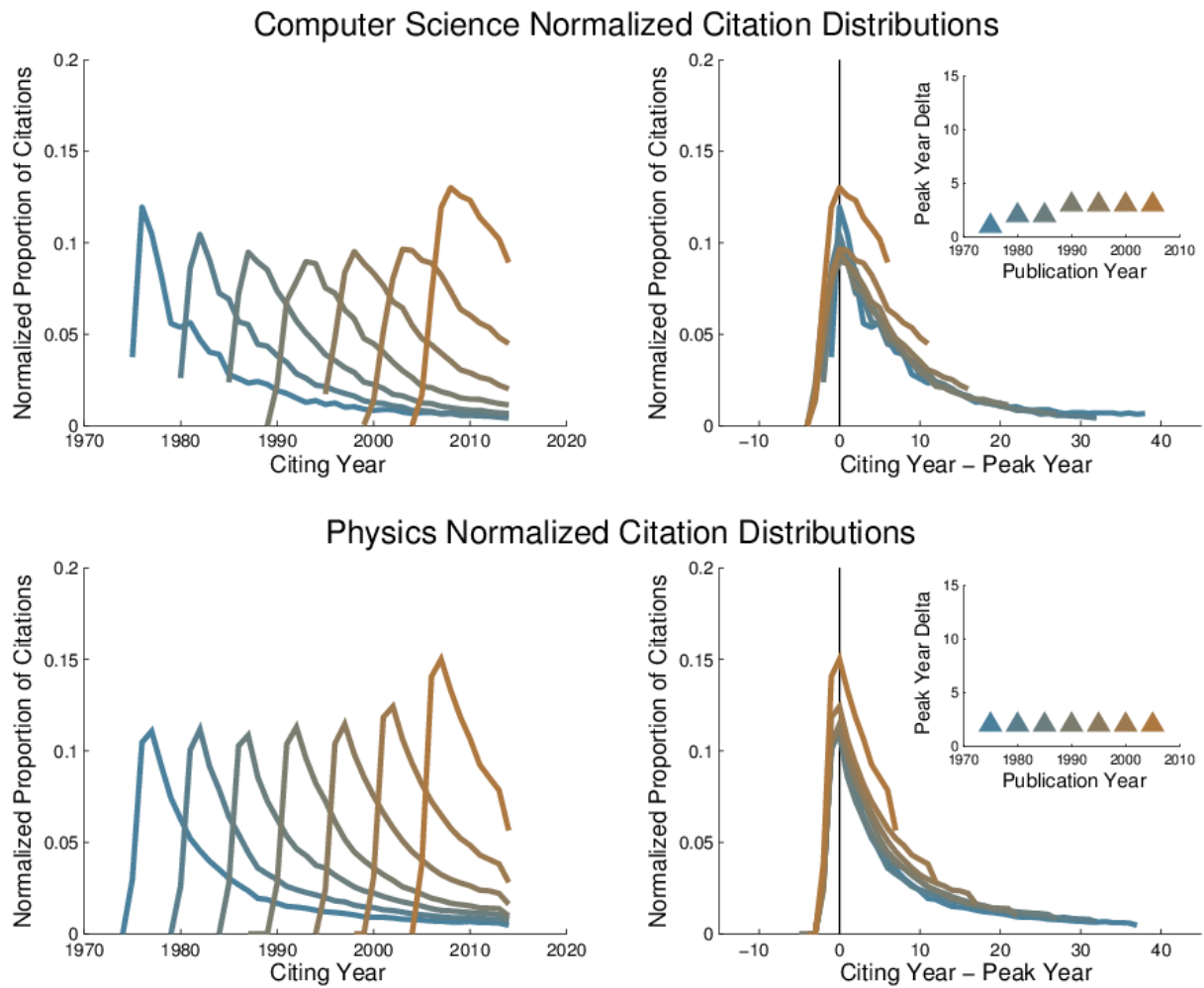

**FIG. 4.** Normalized citation distributions in different years.

Notice that some of the distributions in Figure 3 (and consequently also in Figure 4) seem to cross the horizontal axes. This happens even before the year under scrutiny which actually means citations from the past. A possible explanation may be simply different journal publication delays or that some papers cited conference papers from the same or previous year that appeared as journal articles in later years and the citations were later linked to those future journal articles.

The above analysis focused on the proportion of citations papers published in a particular year receive in each subsequent year. On the other hand, many studies in the past actually considered the proportion of papers that received a specific number of citations over a certain time span. This is in fact the in-degree of the node representing the paper in the underlying citation network. More precisely, let $[y_1, y_2]$ be the time span considered and denote $n_{y_1 y_2}$ to be the number of papers published between years $y_1$ and $y_2$, $n_{y_1 y_2} = \sum_{y \in [y_1, y_2]} n_y$. Furthermore, denote $n_i^y$ to be the number of citations received by *i*-th paper from papers published in year $y$ and $k_i$ the total number of such citations or the in-degree of node *i*, $k_i = \sum_{y \in [y_1, y_2]} n_i^y$. Then, the in-degree distribution of the corresponding citation network is defined as $P_{y_1 y_2}(k) = \sum_{i=1}^{n_{y_1 y_2}} \delta(k_i, k) / n_{y_1 y_2}$, where $\delta$ is the Kronecker delta operator. As first observed by De Solla Price (1965), the tail of the distribution $P_{y_1 y_2}(k)$ follows a power law $\sim k^{-\alpha}$. We refer to $\alpha$ as the power law exponents of citation distributions that are computed using maximum likelihood estimation.

Figure 5 shows the power law exponents $\alpha$ of the distributions of papers' citations in 10-year time spans over the period under investigation. According to the densification law studied by Leskovec et al. (2007), the exponents $\alpha$ should decrease as the network grows. In the left-hand plot of Figure 5 we can indeed observe a moderate decrease in physics, while the exponents are increasing in the case of computer science. The increase starts in 1995 that corresponds to a time span between 1990 and 2000, which is consistent with the change in the citation distributions observed in Figure 3. Thus, the observed change impacts also the structure of citation networks. However, when the distributions of papers' citations are normalized again as in Figure 4, the exponents $\alpha$ decrease in time for both physics and computer science as shown in the right-hand plot of Figure 5. As before, each citation here is counted as $1/n_y$, where $n_y$ in the number of papers in year $y$. The normalized in-degree is thus $\hat{k}_i = \sum_{y \in [y_1, y_2]} n_i^y / n_y$, whereas the normalized in-degree distribution $\hat{P}_{y_1 y_2}(k)$ is defined as above.

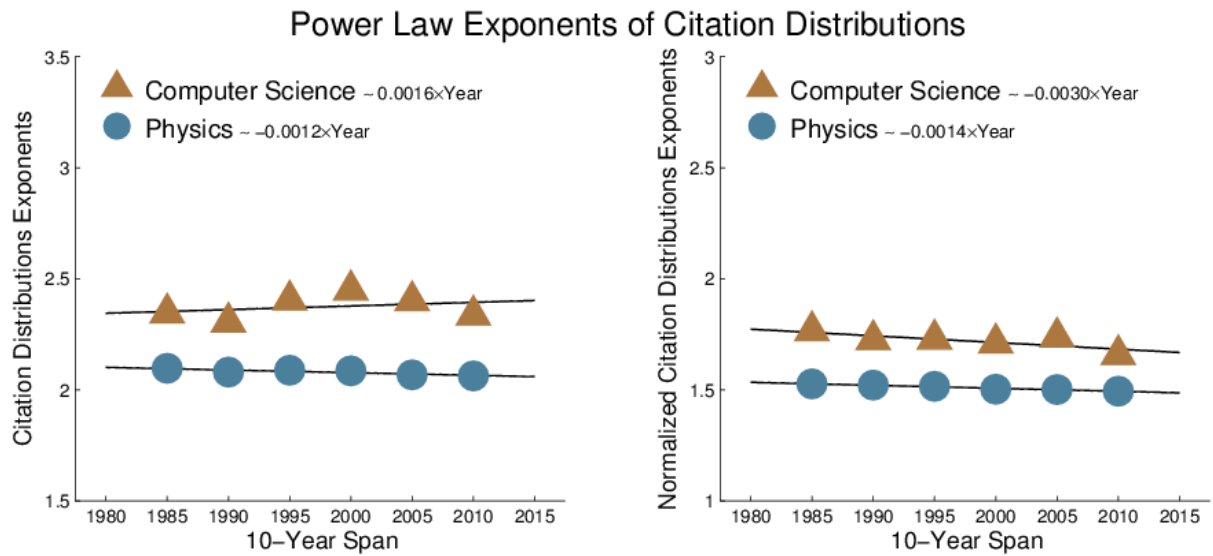

**FIG. 5.** Power law citation exponents in different 10-year spans (straight lines show the least-squares fits to a linear function).

**Conclusion**

There has been an unprecedented publication boom in recent decades resulting in a dramatic increase in the number of research papers based on the expansion of existing journals and conferences and on the emergence of new ones. In this study we were concerned with the question of whether this huge publication growth is also reflected in the way research papers cite each other. We analyzed two large data sets of scientific publications (computer science and physics papers in Web of Science), consisting of half a million and almost two million articles, and identified the major citation trends over the course of time. The main conclusions are that the publication boost in physics is mostly caused by the expansion of the existing publication outlets rather than by the appearance of new ones as it is the case in computer science, and that even though the publication citation peaks of more recent papers seem broader than those of older articles (extremely visible in computer science), these differences are reduced to a minimum if the citation counts are corrected for the growing number of papers. Therefore, the key message of our analysis is that the publication boost in Web of Science journals does indeed alter commonly studied citation distributions, but the overall citation behavior of researchers seems to remain unchanged when citation counts are normalized with respect to the growing number of papers. Note that the unequal coverage of computer science and physics in Web of Science journals in fact allowed for a contrastive comparison in this paper. Future work, however, should investigate this phenomenon for other fields of science and on other data collections as well, to reveal its true origin.

**Acknowledgments**

The authors would like to thank Thomson Reuters for providing bibliographic data and colleagues Ludo Waltman, Nees Jan van Eck, Vladimir Batagelj, and Ján Paralič for helpful suggestions and discussions. Thanks are also due to the reviewers for their insightful comments. For L. Šubelj, this work was supported in part by the Slovenian Research Agency Program No. P2-0359, by the Slovenian Ministry of Education, Science and Sport Grant No. 430-168/2013/91, and by the European Union, European Social Fund. For D. Fiala, this work was supported by the Ministry of Education, Youth and Sports of the Czech Republic within project LO1506 and under grant MSMT MOBILITY 7AMB14SK090.